\documentclass[12pt]{article}
\usepackage{amssymb}
\begin{document}
\title{{\bf Bosonic Reduction of Susy Generalized Harry Dym Equation}}
\author{ Ashok Das$^{a}$ and Ziemowit Popowicz$^{b}$\\
\\
$^{a}$ Department of Physics and Astronomy\\
University of Rochester\\
Rochester, NY 14627 - 0171, USA\\
\\
$^{b}$ Institute of Theoretical Physics\\
University of Wroc\l aw\\
pl. M. Borna 9, 50 -205 Wroc\l aw, Poland\\ }

\maketitle

\begin{center}
{\bf Abstract}
\end{center}

In this paper we construct the two component supersymmetric
generalized Harry Dym equation which is integrable and study various
properties of this model in the bosonic limit. In particular, in the
bosonic limit we obtain a new integrable system which, under a hodograph
transformation, reduces to a coupled three component system. We show
how the Hamiltonian structure transforms under a hodograph
transformation and study the properties of the model under a further
reduction to a two component system. We find a third Hamiltonian
structure for this system (which has been shown earlier to be a
bi-Hamiltonian system) making this a genuinely tri-Hamiltonian
system. The connection of this system to the modified dispersive water
wave equation is clarified. We also study various properties in the
dispersionless limit of our model.

\newpage

\section{Introduction}

The Harry Dym equation \cite{kruskal,maciej} is an important dynamical
system which is
integrable and finds applications in several physical
systems. Together with the Hunter-Zheng equation
\cite{hunter,camasa,brunelli,dai,deformed}, they define a
hierarchy of integrable systems for both positive and negative
integers. This system has been vigorously studied in the past as well
as more recently from various points of view. In particular, the
supersymmetrization of such a system has been discused systematically
in \cite{brunelli1,liu} and a two component generalized Harry Dym equation
has been constructed in \cite{ziemek}.

The construction of supersymmetric integrable systems and the 
understanding of their properties is important for a variety of
reasons \cite{manin,mathieu,ziemek1}. One of the interesting features
lies in the fact that
supersymmetric integrable models can lead to new integrable systems in
the bosonic limit \cite{ziemek1,oevel,delduc}. As we have already
pointed out in connection with
the $N=2$ supersymmetrization of the Harry Dym equation, one of the
supersymmetrizations leads to an interesting new coupled integrable
system  in the bosonic limit \cite{brunelli1}. It is with this aim
that we have chosen 
to construct and study the two component supersymmetric (susy)
generalized Harry Dym equation in this paper. The construction is
quite interesting and leads to many new features including the fact
that in the bosonic limit we obtain new integrable equations.

Our paper is organized as follows. In section {\bf 2}, we recapitulate
briefly the two component generalized Harry Dym equation and some of
its properties. In section {\bf 3}, we construct the two component
supersymmetric Harry Dym equation which is integrable. The Lax
representation as well as the Hamiltonian structures are discussed in
detail in terms of different variables. We construct the hodograph
transformation for the bosonic limit of such a system. In section {\bf
  4}, we address the
question of how a Hamiltonian structure transforms under a hodograph
transformation. Using this, we derive the Hamiltonian structure for
the transformed equation and show that under some
approximation, the system reduces to a three component coupled MKdV
system of equations. A further reduction, in section {\bf 5} takes
this  equation to one
which has been studied earlier and we find a new Hamiltonian structure
for this system which genuinely makes it a tri-Hamiltonian system. The
connection of this system with the modified dispersive water wave
equation is also clarified. In section {\bf 6}, the dispersionless
limit of our system of equations is studied systematically and various
associated properties including polynomial and nonpolynomial charges
are derived. We conclude with a brief summary in section {\bf 7}.

\section{Harry Dym and the Generalized Harry Dym Equations:}

As is well known, the Harry  Dym equation \cite{kruskal} can be
written in  the form 
\begin{equation}
\frac{\partial w}{\partial t} = w^3 w_{xxx},
\end{equation}
where subscripts denote derivatives with respect to the corresponding
variables and the equation can be obtained from the following Lax
representation   
\begin{equation}
\frac{\partial L}{\partial t} = 4\Big [(L)^{\frac{2n+1}{2}}_{\geq 2},L\Big ],
\end{equation}
with $n=1$ and 
\begin{equation}
L= w^2\partial^2. \label{harr}
\end{equation}
Here $\geq 2$ refers to the projection onto the sub-algebra of the pseudo
differential operator $P$, namely,
\begin{equation}
\left(P\right)_{\geq 2} = \sum_{i=2}^{\infty} a_i\partial^{i}.
\end{equation}
The hierarchy of Harry Dym equations is integrable and finds 
applications in various physical examples.

The Harry Dym equation can be generalized to two dynamical variables
in the following way \cite{ziemek}. Let us consider a Lax operator in
the product form
\begin{equation}
L = w^2\partial^2u^2\partial^2,
\end{equation}
where $w,u$ denote two dynamical variables. It is straightforward to
check that the  non-standard Lax representation  
\begin{equation}
\frac{\partial L}{\partial t} = 4\Big [(L)^{3/4}_{\geq 2},L\Big ],
\end{equation}
leads  to the system of coupled equations for $w$ and $u$ of the form 
\begin{eqnarray}
\frac{\partial w}{\partial t} &=& w^3\Big (w^{-1/2}u^{3/2}\Big
)_{xxx}\nonumber\\ 
\frac{\partial u}{\partial t} &=& u^3\Big (u^{-1/2}w^{3/2}\Big
)_{xxx}. \label{row}
\end{eqnarray}
This is the two component generalized Harry  Dym equation which can be
written in the Hamiltonian form as
\begin{equation}
\left(
\begin{array}{c}
w \\ 
\noalign{\vskip 4pt}%
 u
\end{array}
\right)_t = \left(
\begin{array}{cc}
0 & w^3\partial^3 u^3  \\  
\noalign{\vskip 4pt}%
u^3\partial^3 w^3 & 0 \end{array}\right) 
\left (
\begin{array}{c} \frac{\delta H}{\delta w}\\ 
\noalign{\vskip 4pt}%
 \frac{\delta H}{\delta
    u}\end{array} \right),
\end{equation}
where 
\begin{equation}
H =-2 \int \mathrm{d}x\, \Big (wu\Big )^{-\frac{1}{2}}.
\end{equation}
We note that we can rewrite the system of equations (\ref{row}) in a
simpler form in terms of the new variables 
\begin{equation}
u=a e^b \quad , \quad w = ae^{-b},\label{parameterization}
\end{equation}
as
\begin{eqnarray}
a_t &=& a^3 \big ( a_{xxx} + 12a_xb_x^2 +12b_{xx}b_xa \big ),\nonumber\\ 
\noalign{\vskip 4pt}%
b_t &=& -2a^2 \big( (ab)_{xxx} - a_{xxx}b +4b_x^3a
\big).\label{generalized} 
\end{eqnarray}
This equation can be writen in the Hamiltonian form
\begin{equation}
\left(\begin{array}{c}
a\\ 
\noalign{\vskip 4pt}%
 b\end{array}\right)_{t} = \frac{1}{4}\left(\begin{array}{cc}
a^{3}\left(X-X^{\dagger}\right) a^{3} & -a^{3} \left(X +
X^{\dagger}\right) a^{2}\\
\noalign{\vskip 4pt}%
a^{2} \left(X + X^{\dagger}\right) a^{3} & - a^{2}\left(X -
X^{\dagger}\right) a^{2}
\end{array}\right) \left(\begin{array}{c}
\frac{\delta H}{\delta a}\\
\noalign{\vskip 4pt}%
\frac{\delta H}{\delta b}\end{array}\right),
\end{equation}
where the operator $X$ can be identified with
\begin{equation}
X = e^{2b} \partial^{3} e^{-2b},
\end{equation}
with $X^{\dagger}$ representing the Hermitian conjugate and 
\begin{equation}
H = -2 \int \mathrm{d}x\, a^{-1}.
\end{equation}

\section{Supersymmetric Generalized Harry Dym Equation:}

The supersymmetric extensions of the Harry Dym equation have already been
constructed in \cite{brunelli1}. Let us briefly recapitulate some of
the features of the $N=2$
extended supersymmetric Harry Dym equation which is most useful from
the point of view of constructing the generalized supersymmetric Harry
Dym equation. In this case the bosonic dynamical variable is 
generalized to a bosonic $N=2$ superfield 
\begin{equation}
W = w_0 + \theta_1 \chi_1 + \theta_2\chi_2 + \theta_1\theta_2 w_1,
\end{equation} 
where $w_1$ is the original bosonic dynamical variable while $ w_0 $
is the new bosonic dynamical variable and  
$\chi_i, i=1,2$ are the new fermionic dynamical variables necessary
for $N=2$ supersymmetry. We note that $\theta_{i}, i=1,2$ represent
the two Grassmann coordinates of the $N=2$ superspace. We can define
the two supercovariant derivatives on this space as
\begin{equation}
D_1 = \frac{\partial }{\partial \theta_1} +
\theta_1\frac{\partial}{\partial x} \quad , \quad  
D_2 = \frac{\partial }{\partial \theta_2} +
\theta_2\frac{\partial}{\partial x},
\end{equation}
which satisfy
\begin{equation}
\left\{D_{1}, D_{2}\right\} = 0, \quad D_{1}^{2} = D_{2}^{2} =
\partial.
\end{equation}
 
In \cite{brunelli1} we showed that it is possible to construct four
different $N=2$ 
supersymmetric Lax operators that lead to consistent 
equations. Namely, the general Lax operator with the $N=2$ superfield
$T$ 
\begin{eqnarray}
L &=& T^{-1}\partial^2 +k_1(D_1T^{-1})\partial
+k_1(D_2T^{-1})\partial  +  \nonumber\\ 
&& (k_2(D_1 D_2 T)T^{-2} + k_3(D_1 T)(D_2 T)T^{-3})D_1 D_2,\label{Lsusy}  
\end{eqnarray}
leads to a consistent non-standard Lax representation 
\begin{equation}
\frac{\partial L}{\partial t} = 4\Big [(L)^{\frac{3}{2}}_{\geq 2},L\Big
],\label{susyLequation}
\end{equation}
only for $ k_1=k_2=k_3=0 $ or $ k_1=k_2=-\frac{k_3}{2} =1 $ or $
k_1=\frac{1}{2} , k_2=k_3=0 $  or  
$ k_1=k_2=\frac{1}{2} , k_3=\frac{3}{4}$. 
We will consider here only the last case which allows us to carry out 
the construction of the  
two component supersymmetric generalized Harry Dym equation. 

Let us note (which has been pointed out in \cite{brunelli1} as well)
that in this
particular case, the Lax operator (\ref{Lsusy}) can be written as 
\begin{equation}
L=-(W D_1 D_2)^2,\label{N2susyL}
\end{equation}
where $W$ is related to $T$ and the Lax equation (\ref{susyLequation})
leads  to the dynamical equations  
\begin{eqnarray} 
W_t &=& \frac{1}{8} \Big( 6(D_1 D_2 W)(D_1 D_2 W_x)W^2 -2W_{xxx}W^3
\nonumber\\   
&&\;\; + 3(D_1 W_{xx})(D_1W)W^2 + 3(D_2 W_{xx})(D_2W)W^2
\Big). \label{susyhd} 
\end{eqnarray} 
This equation can be written in the Hamiltonian form
\begin{equation}
W_t={\cal D}_1 \frac{\delta H_0}{\delta W},
\end{equation}
where 
\begin{equation}
{\cal D}_{1} = - \frac{1}{2} W^{2} D_{1} D_{2} \partial W^{2},\quad
H_0 = -\frac{1}{4} \int \mathrm{d}x\,\mathrm{d}^{2}\theta\, (D_2
W)(D_1 W)W^{-1},
\end{equation}
with $\mathrm{d}^{2}\theta =
\mathrm{d}\theta_{1}\,\mathrm{d}\theta_{2}$. 

In order to construct the two component supersymmetric generalized
Harry Dym 
equation, let us consider the two differential operators on the $N=2$
superspace  
\begin{equation}
L_1 = F D_1 D_2,\quad L_2 =  G D_1 D_2,
\end{equation}
each of which has the form of the square root of (\ref{N2susyL}). Here
$F,G$ denote two bosonic superfields on the $N=2$ superspace. Let us
next construct a new Lax operator as the product
\begin{equation}
L = - L_1L_2.
\end{equation}
It can be checked that the two component susy generalized Harry Dym
equation  follows from the non-standard Lax representation 
\begin{equation}
\frac{\partial L}{\partial t} = 4[ L^{\frac{3}{2}}_{\geq 2},L ]
\end{equation}
In fact, the equations are derived in a much simpler manner with a
parameterization of the form (\ref{parameterization}), namely,
\begin{equation}
F = e^W U,\quad  G=e^{-W} U.
\end{equation}
With this parametrization, the dynamical equations can be written in
the Hamiltonian form
\begin{equation}
\left(\begin{array}{c}U\\
\noalign{\vskip 4pt}%
 W \end{array}\right )_t = \frac{1}{4}
\left(\begin{array}{cc} U^2 \left(X - X^{\dagger}\right) U^2 & -U^2
\left(X + X^{\dagger}\right) U  \\
\noalign{\vskip 4pt}%
 U \left(X + X^{\dagger}\right) U^2 &  -U \left(X - X^{\dagger}\right)
U \end{array} \right ) \left(\begin{array}{c}
\frac{\delta H_{2}}{\delta U}\\
\noalign{\vskip 4pt}%
\frac{\delta H_{2}}{\delta W}
\end{array}\right),\label{equation}
\end{equation}
where    $ X =e^W D_1 D_2 \partial e^{-W}$ and  
\begin{equation} 
H_2= {\rm Tr}\Big (L^{\frac{1}{2}}\Big ) =  \int
\mathrm{d}x\,\mathrm{d}^{2}\theta\left[(D_2 W)(D_1 W)U - \frac{(D_2
U)(D_1 U)}{U}\right].
\end{equation}
It is  also possible to write these equations in a still simpler form
by introducing the variables
\begin{equation} 
U = \frac{1}{\sqrt{fg}}, \quad  W = \frac{1}{2} \left(\ln g - \ln f\right),
\end{equation}
in terms of which we have 
\begin{equation}
\left(\begin{array}{c}f\\
\noalign{\vskip 4pt}%
 g \end{array}\right )_t =
\left(\begin{array}{cc} 0  &  Y \\
\noalign{\vskip 4pt}%
- Y^{\dagger}  & 0 \end{array} \right ) \left(\begin{array}{c}
\frac{\delta H_2}{\delta f}\\
\noalign{\vskip 4pt}%
\frac{\delta H_{2}}{\delta g}\end{array}\right),\label{hs}
\end{equation}
where
\[
Y = D_{1}D_{2}\partial,\quad H_{2} = - 4 \int
\mathrm{d}x\,\mathrm{d}^{2}\theta\,
\frac{1}{\sqrt{f}}\,D_{1}D_{2}\left(\frac{1}{\sqrt{g}}\right).
\]
It is now obvious that the Hamiltonian structure in (\ref{hs})
satisfies Jacobi identity.
We will, however, continue to work with the first representation of
the  Hamiltonian equations (\ref{equation}). This system of equations
is integrable and the conserved charges can be obtained from the
traces of the Lax operator in the standard manner.

In the bosonic limit where we can restrict the superfields and the
Hamiltonian  to the forms  
\begin{eqnarray}
W & = & w_0 + \theta_1\theta_2 w_1 , \quad U = u_0 + \theta_1\theta_2
 u_1 ,\nonumber\\
\noalign{\vskip 4pt}%
H_2 & = & \int \mathrm{d}x\,\left(w_{0x}^2u_0+w_{1}^2u_0
-\frac{u_{0x}^2+u_1^2}{u_0}\right),
\end{eqnarray}
equation (\ref{equation}) leads to a system of four interacting
equations of the form
\begin{eqnarray}
w_{0t} &=&
\frac{1}{2}\left(2w_{0xxx}u_{0}^3+6w_{0xx}u_{0x}u_{0}^2-w_{0x}^3u_{0}^3 
+3w_{0x}u_{0x}^2u_{0}\right.  \nonumber \\
&& \quad \left. +3w_{0x}u_1^2u_{0} -3w_1^2w_{0x}u_{0}^3\right), \nonumber\\ 
w_{1t} &=&
\frac{1}{2}\left(-6w_{0xx}u_{1x}u_{0}^2-3w_{0x}^3u_1u_{0}^2
-6w_{0x}u_{1xx}u_{0}^2 -12w_{0x}u_{1x}u_{0x}u_{0}\right. \nonumber\\
&& \quad + 3w_{0x}u_1^3 -6w_{0x}u_1u_{0xx}u_{0} +3w_{0x}u_1u_{0x}^2
+2w_{1xxx}u_{0}^3 \nonumber  \\
&& \quad + 12w_{1xx}u_{0x}u_{0}^2 +12w_{1x}u_{0xx}u_{0}^2
+12w_{1x}u_{0x}^2u_{0}  +6w_{1x}w_{0x}^2u_{0}^3 \nonumber\\
&& \quad + 6w_{1x}w_1^2u_{0}^3 +6w_1^3u_{0x}u_{0}^2
-3w_1^2w_{0x}u_1u_{0}^2 +6w_1u_{0xxx}u_{0}^2 \nonumber\\
&& \quad \left.+ 12w_1u_{0xx}u_{0x}u_{0} -6w_1u_{1x}u_1u_{0}
+6w_1w_{0xx}w_{0x}u_{0}^3 + 12w_1w_{0x}^2u_{0x}u_{0}^2\right), \nonumber\\
u_{0t} &=&
\left(u_{0xxx}u_{0}^3-3u_{1x}u_1u_{0}^2+3w_{1x}w_1u_{0}^4
+3w_1^2u_{0x}u_{0}^3\right),\nonumber\\ 
u_{1t} &=&
\frac{1}{2}\left(2u_{1xxx}u_{0}^3+6u_{1xx}u_{0x}u_{0}^2+6u_{1x}u_{0xx}u_{0}^2
-12u_{1x}u_1^2u_{0}\right.\nonumber\\ 
&&\quad +6u_1u_{0xxx}u_{0}^2 +6w_{0xx}w_{0x}u_1u_{0}^3
+6w_{0x}^2u_{1x}u_{0}^3 +6w_{0x}^2u_1u_{0x}u_{0}^2\nonumber\\
&& \quad
-6w_{1xx}w_{0x}u_{0}^4-6w_{1x}w_{0xx}u_{0}^4-24w_{1x}w_{0x}u_{0x}u_{0}^3
+12w_{1x}w_1u_1u_{0}^3 \nonumber\\ 
&& \quad
-3w_1^3w_{0x}u_{0}^4+12w_1^2u_1u_{0x}u_{0}^2-6w_1w_{0xx}u_{0x}u_{0}^3
-3w_1w_{0x}^3u_{0}^4\nonumber\\
&&\quad  \left.-12w_1w_{0x}u_{0xx}u_{0}^3-9w_1w_{0x}u_{0x}^2u_{0}^2
+3w_1w_{0x}u_1^2u_{0}^2\right).\label{equation1}
\end{eqnarray}
When $w_{0}=w_{1}=0$, this system of equations reduces to the bosonic
limit of the $N=2$ susy Harry Dym equation (\ref{susyhd}). On the
other hand, when we set $w_{0}=u_{1}=0$, and identify
$u_{0}=a,w_{1}=2b_{x}$ the system of equations goes over to the
generalized Harry Dym 
equation (\ref{generalized}) (except for
an overall factor in the second equation which we are unable to get
rid of by any scaling).

Let us next study the behavior of this system of equations under a
hodograph transformation \cite{ibragimov} of the variables
$(x, t)$ to $(y,\tau)$ defined through  
\begin{equation}
x=p_0(y,\tau), \quad \tau = t,
\end{equation}
so that we have 
\begin{eqnarray}
\frac{\partial }{\partial x} &=& \frac{\partial y }{\partial
  x}\frac{\partial }{\partial y} = \frac{1}{p_{0y}}\frac{\partial
  }{\partial y}, \nonumber\\  
\frac{\partial }{\partial \tau} &=& \frac{\partial}{\partial t} +
  p_{0\tau} \frac{\partial}{\partial x} = \frac{\partial }{\partial t} +
  \frac{p_{0\tau}}{p_{0y}}\frac{\partial}{\partial
  y}. \label{hodograph1} 
\end{eqnarray}
Defining the variables 
\begin{equation}
w_0=q_0,\quad w_1=q_1, \quad u_0 =p_{0y},\quad u_1
=p_1,\label{hodograph2} 
\end{equation}
the system of equations (\ref{equation1}) goes over to 
\begin{eqnarray}
p_{0\tau} &=&\frac{1}{2}
\left(2p_{0yyy}-3p_{0yy}^2p_{0y}^{-1}-3p_1^2p_{0y}+3q_{1}^2p_{0y}^3\right),
\nonumber\\ 
p_{1\tau} &=&\frac{1}{2}
\left(2p_{1yyy}+6p_{1y}p_{0yyy}p_{0y}^{-1}-9p_{1y}p_{0yy}^2p_{0y}^{-2}
-15p_{1y}p_{1}^2\right.\nonumber\\  
&&
\quad +6p_{1}p_{0yyyy}p_{0y}^{-1}-24p_{1}p_{0yyy}p_{0yy}p_{0y}^{-2}
+18p_{1}p_{0yy}^3p_{0y}^{-3}+6q_{0yy}q_{0y}p_{1}\nonumber\\  
&&
\quad +6q_{0y}^2p_{1y}-6q_{1yy}q_{0y}p_{0y}-6q_{1y}q_{0yy}p_{0y}
-12q_{1y}q_{0y}p_{0yy}\nonumber\\  
&&
\quad +12q_{1y}q_1p_1p_{0y}^2-3q_{1}^3q_{0y}p_{0y}^3+3q_{1}^2p_{1y}p_{0y}^2
+12q_{1}^2p_1p_{0yy}p_{0y}\nonumber\\  
&&\quad
-6q_1q_{0yy}p_{0yy}-3q_{1}q_{0y}^3p_{0y}-12q_1q_{0y}p_{0yyy}
+9q_1q_{0y}p_{0yy}^2p_{0y}^{-1}\nonumber\\
\noalign{\vskip 4pt}%
 & & \quad \left.+3q_1q_{0y}p_{1}^2p_{0y}\right),
\nonumber\\ 
q_{0\tau} &=& \frac{1}{2}\left(2q_{0yyy}-q_{0y}^3\right),\nonumber\\ 
q_{1\tau} &=&\frac{1}{2}
\left(-6q_{0yy}p_{1y}p_{0y}^{-1}-3q_{0y}^3p_{1}p_{0y}^{-1}
-6q_{0y}p_{1yy}p_{0y}^{-1}\right.\nonumber\\  
&&
\quad +3q_{0y}p_{1}^3p_{0y}^{-1}-6q_{0y}p_{1}p_{0yyy}p_{0y}^{-2}
+9q_{0y}p_{1}p_{0yy}^2p_{0y}^{-3}+2q_{1yyy}\nonumber\\  
&&
\quad +6q_{1yy}p_{0yy}p_{0y}^{-1}+12q_{1y}p_{0yyy}p_{0y}^{-1}
-9q_{1y}p_{0yy}^2p_{0y}^{-2}-3q_{1y}p_{1}^2\nonumber\\  
&&
\quad +6q_{1y}q_{0y}^2+9q_{1y}q_{1}^2p_{0y}^2+6q_{1}^3p_{0yy}p_{0y}
-3q_{1}^2q_{0y}p_{1}p_{0y}\nonumber\\  
&&
\quad +6q_{1}p_{0yyyy}p_{0y}^{-1}-12q_{1}p_{0yyy}p_{0yy}p_{0y}^{-2}
+6q_{1}p_{0yy}^3p_{0y}^{-3}-6q_{1}p_{1y}p_{1}\nonumber\\  
&&
\quad\left.
+6q_{1}q_{0yy}q_{0y}+6q_{1}q_{0y}^2p_{0yy}p_{0y}^{-1}\right).\label{equation2}
\end{eqnarray}

We note from (\ref{equation2}) that the equation for the variable
$q_{0}$ is decoupled. As a result,  we can set $q_0=0$ for simplicity. 
The Hamiltonian structure for the system of equation involving 
$w_1,u_0,u_1$ can be obtained from the Dirac reduction of the   
bosonic limit of the Hamiltonian operator in (\ref{equation}).
In the simple case of two variables, for example, the Dirac reduction
works as follows. Let us assume the equations of motion to have the
form   
\begin{equation}
\left( \begin{array}{c}u\\ 
\noalign{\vskip 4pt}%
v \end{array}\right )_t = \left( \begin{array}{cc}
P_{uu} & P_{uv} \\ 
\noalign{\vskip 4pt}%
P_{vu} & P_{vv} \end{array} \right ) \left(\begin{array}{c}
\frac{\delta H}{\delta u}\\
\noalign{\vskip 4pt}%
\frac{\delta H}{\delta v}
\end{array}\right),
\end{equation}
where $ u,v $ denote the two dynamical variables with $ P_{uu},
P_{uv}, P_{vu}, P_{vv} $ representing the elements of the Hamiltonian  
operator. If we can set $v=0$, then it can be verified directly 
that \cite{maciej,brunelli2} the Hamiltonian structure reduces to
\begin{equation}
u_t = \Big ( P_{uu} - P_{uv} P_{vv}^{-1} P_{vu} \Big ) \left.\frac{\delta
  H}{\delta u}\right|_{v=0},
\end{equation}
for the reduced system.

For the three component system under study in (\ref{equation1}) (with
$w_{0}=0$), the Dirac reduction leads to 
\begin{equation}
\left(\begin{array}{c}u_0\\ 
\noalign{\vskip 4pt}%
u_1 \\
\noalign{\vskip 4pt}%
w_1 \end{array}\right )_t = {\cal D} \left(\begin{array}{c}
\frac{\delta H_{2}}{\delta u_{0}}\\
\noalign{\vskip 4pt}%
\frac{\delta H_{2}}{\delta u_{1}}\\
\noalign{\vskip 4pt}%
\frac{\delta H_{2}}{\delta w_{1}}
\end{array}\right),
\end{equation}
where the matrix Hamiltonian operator ${\cal D}$ has the elements
\begin{eqnarray}
{\cal D}^{(1,1)} &=& \partial u_0^4 + u_0^4 \partial,\nonumber\\
{\cal D}^{(1,2)} &=& 4u_{1y}u_0^3 + 4u_1u_{0y}u_0^2 +
4u_1u_0^3\partial, \nonumber\\
{\cal D}^{(1,3)} &=& 2w_{1y}u_0^3 + 2w_{1}u_{0y}u_0^2+
2w_1u_0^3\partial, \nonumber\\ 
{\cal D}^{(2,2)} &=& -\partial^3 u_0^4 - u_0^4 \partial^3 - \partial
(12u_{0y}^2u_0^2 -4u_1^2u_0^2) - (12u_{0y}^2u_0^2 - 4u_1^2u_0^2)
\partial,\nonumber\\  
{\cal D}^{(2,3)} &=&  4w_{1y}u_1u_0^2 + 4w_1u_1u_{0y}u_0
+4w_1u_1u_0^2\partial, \nonumber\\ 
{\cal D}^{(3,3)} &=& \partial^2 u_0^3 + u_0^2\partial^3  + \partial
(w_1^2u_0^2-3u_{0y}^2)+(w_1^2u_0^2-3u_{0y}^2) \partial,  
\end{eqnarray}
and  
\begin{equation}
H_2= \int \mathrm{d}x\,\left(w_{1}^2u_0
-\frac{u_{0x}^2+u_1^2}{u_0}\right). \label{hamiltonian}
\end{equation}

\section{Transformation of the Hamiltonian Structure under the
  Hodograph Transformation:}

In this section, we will discuss how the Hamiltonian structure
transforms under a hodograph transformation. Namely, we already have
the Hamiltonian structure for the three component equation in
(\ref{equation1}) with $w_{0}=0$. We would like to determine the
Hamiltonian structure for the three component equation in
(\ref{equation2}) with $q_{0}=0$ which is obtained from
(\ref{equation1}) under a hodograph transformation.
We note that the Hamiltonian (\ref{hamiltonian}) in terms of the three
variables $ p_0, p_1$ and $q_1$ takes the form 
\begin{equation} 
H = \int \mathrm{d}y\,\left(-\frac{p_{0yy}}{p_{0y}^2} - p_1^2
+q_1^2p_{0y}^2\right). 
\end{equation}
Furthermore, from the definition of the hodograph transformations in
(\ref{hodograph1}) and (\ref{hodograph2}), we obtain
\begin{equation}
\left(\begin{array}{c}u_0\\ u_1 \\w_1 \end{array}\right )_t = 
S^{-1} \left(\begin{array}{c}p_0\\ p_1 \\q_1 \end{array}\right )_{\tau},
\end{equation}
where $S$ has the form
\begin{equation}
S =\left(\begin{array}{ccc}
p_{0y}\partial^{-1}p_{0y}^{-1}  & 0 & 0  \\
\noalign{\vskip 4pt}%
p_{1y}\partial^{-1}p_{0y}^{-1} &  1  &  0 \\
\noalign{\vskip 4pt}%
q_{1y}\partial^{-1}p_{0y}^{-1} & 0  & 1  \end{array}\right).
\end{equation}
The gradient of the Hamiltonian transforms under the hodograph
transformation as 
\begin{equation}
\left(\begin{array}{c} \frac{\delta H}{\delta u_{0}} \\
\noalign{\vskip 4pt}%
 \frac{\delta
    H}{\delta u_{1}} \\ 
\noalign{\vskip 4pt}%
\frac{\delta H}{\delta w_{1}} \end{array}\right ) = 
 K \left(\begin{array}{c}\frac{\delta H}{\delta p_0} \\
\noalign{\vskip 4pt}%
 \frac{\delta
    H}{\delta p_1} \\ 
\noalign{\vskip 4pt}%
 \frac{\delta H}{\delta q_1} \end{array}\right),
\end{equation}
where 
\begin{equation}
K =\left(\begin{array}{ccc}
-p_{0y}^{-2}\partial^{-1}p_{0y} &  -p_{0y}^{-2}\partial^{-1}p_{1y} &
-p_{0y}^{-2}\partial^{-1}q_{1y}  \\ 
\noalign{\vskip 4pt}%
0 & p_{0y}^{-1} & 0  \\
\noalign{\vskip 4pt}%
0  & 0 & p_{0y}^{-1}  \end{array}\right ) =p_{0y}^{-1}S^{\dagger}
\end{equation}
It follows now that
\begin{equation}
\left(\begin{array}{c}p_0\\ 
\noalign{\vskip 4pt}%
p_1 \\
\noalign{\vskip 4pt}%
q_1 \end{array}\right )_{\tau}= S {\cal
     D}\left(\begin{array}{c}
\frac{\delta H_{2}}{\delta u_{0}}\\
\noalign{\vskip 4pt}%
\frac{\delta H_{2}}{\delta u_{1}}\\
\noalign{\vskip 4pt}%
\frac{\delta H_{2}}{\delta w_{1}}
\end{array}\right) 
 = S{\cal D}K \left(\begin{array}{c}
\frac{\delta H_{2}}{\delta p_{0}}\\
\noalign{\vskip 4pt}%
\frac{\delta H_{2}}{\delta p_{1}}\\
\noalign{\vskip 4pt}%
\frac{\delta H_{2}}{\delta q_{1}}
\end{array}\right) = \widetilde{\cal D} \left(\begin{array}{c}
\frac{\delta H_{2}}{\delta p_{0}}\\
\noalign{\vskip 4pt}%
\frac{\delta H_{2}}{\delta p_{1}}\\
\noalign{\vskip 4pt}%
\frac{\delta H_{2}}{\delta q_{1}}
\end{array}\right).
\end{equation}
The form of the transformed Hamiltonian structure $\widetilde{\cal D}$ can now
be determined easily to have the following elements
\begin{eqnarray}
\widetilde{\cal D}^{(1,1)} &=&  - 2p_{0y}\partial^{-1}p_{0y},\nonumber\\
\noalign{\vskip 4pt}%
\widetilde{\cal D}^{(1,2)} &=&  -2p_{0y}\partial^{-1}p_{1y} +
4p_{1}p_{0y}, \nonumber\\ 
\widetilde{\cal D}^{(1,3)} &=&  -2p_{0y}\partial^{-1}q_{1y} + 2q_1p_{0y},
\nonumber\\ 
\widetilde{\cal D}^{(2,2)} &=& -2(\partial p_{0yyy}p_{0y}^{-1} +
p_{0yyy}p_{0y}^{-1}\partial ) + 6(\partial p_{0yy}^2p_{0y}^{-2} 
+p_{0yy}^2p_{0y}^{-2}\partial )\nonumber\\ 
&&\quad   -2p_{1y}\partial^{-1}p_{1y} +4(\partial p_1^2+p_1^2\partial)
-2\partial^3,\nonumber\\  
\widetilde{\cal D}^{(2,3)} &=& -2p_{1y}\partial^{-1}q_{1y} + 2q_1p_{1y}
+4q_1p_1\partial),\nonumber\\  
\widetilde{\cal D}^{(3,3)} &=& - 2q_{1y}\partial^{-1} q_{1y}+\partial^2
p_{0x}^{-2}\partial + \partial p_{0x}^{-2}\partial^2 +  
\partial  q_1^2 + q_1^2 \partial.   
\end{eqnarray}

We can further simplify the equations (\ref{equation2}) defining the
new variables 
\begin{equation}
f = \frac{ p_{0yy} }{ p_{0y} }, \quad g = p_1, \quad h = q_1p_{0y}, 
\end{equation}
which leads to the equations
\begin{eqnarray}
f_{\tau} &= & \Big ( 2f_{yy} - f^3-6g_{y}g-3g^2f+6h_{y}h +3h^2f \Big
)_y,\nonumber \\
g_{\tau} &=& \Big  ( 2g_{yy} - 5g^3+6gf_{y}-3gf^2+3h^2g  \Big )_y+
6h_{y}hg, \nonumber\\  
h_{\tau} &=& \Big  (2h_{yy} + 5h^3+6hf_{y}-3hf^2-3hg^2 \Big )_y -6
hg_{y}g.\label{equation3} 
\end{eqnarray}
The Hamiltonian structure $\widetilde{\cal D}$, in this case,
transforms to 
\begin{eqnarray}
\widetilde{\cal D}^{(1,1)} &=&    - 2f_y\partial^{-1}f_y +\partial f^2 + f^2
\partial - 2\partial^3,\nonumber\\
\noalign{\vskip 4pt}%
\widetilde{\cal D}^{(1,2)} &=&  -2f_y\partial^{-1}g_y +
2\partial(\partial g + 
g\partial ) + 2\partial gf +2g\partial f,\nonumber\\
\noalign{\vskip 4pt}%
\widetilde{\cal D}^{(1,3)} &=&  -2f_y\partial^{-1}h_y + 2\partial
(\partial h + 
h\partial ) + 2\partial hf + 2h\partial f,\nonumber\\  
\noalign{\vskip 4pt}%
\widetilde{\cal D}^{(2,2)} &=& -2g_y\partial^{-1}g_y  +
\partial(f^2+4g^2-2f_y) + 
(f^2+4g^2-2f_y) \partial - 2\partial^3,\nonumber\\
\noalign{\vskip 4pt}%
\widetilde{\cal D}^{(2,3)} &=& -2g_y\partial^{-1}h_y + 4(\partial hg + hg
\partial), \label{structure}\\  
\noalign{\vskip 4pt}%
\widetilde{\cal D}^{(3,3)} &=& -2h_y\partial^{-1}h_y +\partial
(4h^2-f^2+2f_y) + (4h^2-f^2+2f_y) \partial + 2\partial^3, \nonumber
\end{eqnarray}
and the Hamiltonian in the new variables has the form 
\[H= \int \mathrm{d}y\,\left(- f^2 - g^2 + h^2\right).
\]
All the operators
$\partial$ in the expression (\ref{structure}) refer to derivative
operators with
respect to the variable $y$. We note that $H_0= \int
\mathrm{d}y\, f$ is a conserved quantity which defines the Casimir of
the Hamiltonian structure  $\widetilde{\cal D}$ in (\ref{structure})
in the sense that it annihilates the gradient of the Hamiltonian
\cite{brunelli}. 

\section{Connection with the Modified Dispersive Water Wave Equation:}

It is well known that the Harry Dym equation can be transformed to the
MKdV equation under a hodograph transformation
\cite{ibragimov}. Similarly,  the
generalized Harry Dym equation can be transformed to two coupled MKdV
systems under a hodograph transformation \cite{sakovich}. Such two
component coupled
MKdV systems have already been classified. In this section, we will
study the connection of our system of equations with other equations.  
Let us note that if we set $h=0$ in (\ref{equation2}) (which would
correspond to setting $W=0$ for the supersymmetric generalized  
Harry Dym equation) the system of equations takes the form 
\begin{eqnarray}
f_{\tau} &= & \Big ( 2f_{yy} - f^3-6g_{y}g-3g^2f \Big )_y \nonumber\\ 
g_{\tau} &=& \Big  ( 2g_{yy} - 5g^3+6gf_{y}-3gf^2  \Big )_y. \label{mkdv}
\end{eqnarray}
This is a system of two coupled MKdV equations and 
under  this reduction the Hamiltonian structure can be obtained from 
the  Dirac reduction of the operator $\widetilde{\cal D}$ as discussed
earlier and we obtain the two by two matrix $\widetilde{\cal D}$ with
elements 
\begin{eqnarray}
\widetilde{\cal D}^{(1,1)} &=&    - 2f_y\partial^{-1}f_y +\partial f^2 + f^2
\partial - 2\partial^3, \nonumber\\
\noalign{\vskip 4pt}%
\widetilde{\cal D}^{(1,2)} &=&  -2f_y\partial^{-1}g_y +
2\partial(\partial g + 
g\partial ) + 2\partial gf +2g\partial f, \label{fgstructure}\\  
\widetilde{\cal D}^{(2,2)} &=& -2g_y\partial^{-1}g_y  +
\partial(f^2+4g^2-2f_y) +
(f^2+4g^2-2f_y) \partial - 2\partial^3.\nonumber
\end{eqnarray}
We note that, as in the previous case, $ \int \mathrm{d}y\,f$ is conserved
and defines the Casimir of the Hamiltonian structure. However, in the
present case, $\tilde{H} = \int \mathrm{d}y\, g$ is also conserved and
can be used to construct a new system of equations
\begin{equation}
\left(\begin{array}{c} f\\ 
\noalign{\vskip 4pt}%
g \end{array}\right )_{\tau} = 
\widetilde{\cal D}\left(\begin{array}{c}
\frac{\delta \tilde{H}}{\delta f}\\
\noalign{\vskip 4pt}%
\frac{\delta \tilde{H}}{\delta g}
\end{array}\right) = \left(\begin{array}{c}2(g_y+gf)_y\\
  (-2f_y+3g^2+f^2)_y  \end{array}\right ).\label{new} 
\end{equation}
Under a change of  variables
\begin{equation}
f=(a+b), \quad g=i(a-b),\label{change}
\end{equation} 
the system of equations (\ref{mkdv}) takes the form
\begin{eqnarray}
a_{\tau} &=& (a_{yy}+3aa_y-3ba_y+a^3-6a^2b+3ab^2)_y,\nonumber\\
\noalign{\vskip 4pt}%
b_{\tau} &=& (b_{yy}+3bb_y-3ab_y+b^3-6b^2a+3ba^2)_y,\label{mkdv'} 
\end{eqnarray}
which has been studied earlier by Sakovich \cite{sakovich} and
Foursov \cite{foursov}.
On the other hand, under this change of variables (\ref{change}), the
new system of equations (\ref{new}) has the form 
\begin{eqnarray}
a_{\tau} &=&2i(a_{y}+a^2 -2ab)_y, \nonumber\\
\noalign{\vskip 4pt}%
b_{\tau} &=&2i(-b_{y}-b^2+2ab)_y.\label{new'}
\end{eqnarray}
Theses constitute the system of coupled Burgers equations
\cite{maciej} and in
the following we will neglect the overall factor $2i$. 

Both of the system of equations (\ref{mkdv'}) and (\ref{new'}) are
Hamiltonian with the Hamiltonian structure (it is the structure
(\ref{fgstructure}) in the new variables)
\begin{eqnarray}
\widetilde{\cal D}_0^{(1,1)} &=&    - 2a_y\partial^{-1}a_y
+\partial(a_y+2a^2-2ab) 
+ (a_y+2a^2-2ab)\partial,\nonumber\\ 
\noalign{\vskip 4pt}%
\widetilde{\cal D}_0^{(1,2)} &=&  -2a_y\partial^{-1}b_y -\partial^3
-\partial 
(\partial a -2ab + a^2+  a\partial) + \nonumber\\
\noalign{\vskip 4pt}%
&& (\partial b +2ab -b^2+ b\partial) \partial + 2b\partial
a,\nonumber\\
\noalign{\vskip 4pt}%
{\cal D}_0^{(2,2)} &=&    - 2b_y\partial^{-1}b_y +\partial(b_y+2b^2-2ab)
+ (b_y+2b^2-2ab)\partial.\label{structure0} 
\end{eqnarray}
It is already known that the systems of equations (\ref{mkdv'}) and
(\ref{new'}) are bi-Hamiltonian \cite{maciej,sakovich1} with
Hamiltonian structures $\widetilde{\cal D}_{1},\widetilde{\cal D}_{2}$
which we describe below. 
However we find that, in fact, both these systems of equations are 
tri-Hamiltonian systems much like the two boson equation. For example,
we can write the system of equations (\ref{new'}) as
\begin{equation}
\left(\begin{array}{c}a \\ 
\noalign{\vskip 4pt}%
b \end{array}\right )_{\tau}=\widetilde{\cal D}_0
\left(\begin{array}{c}
\frac{\delta \tilde{H}_{0}}{\delta a}\\
\noalign{\vskip 4pt}%
\frac{\delta \tilde{H}_{0}}{\delta b}
\end{array}\right) = 
\widetilde{\cal D}_1 \left(\begin{array}{c}
\frac{\delta \tilde{H}_{1}}{\delta a}\\
\noalign{\vskip 4pt}%
\frac{\delta \tilde{H}_{1}}{\delta b}
\end{array}\right) = \widetilde{\cal D}_2 \left(\begin{array}{c}
\frac{\delta \tilde{H}_{2}}{\delta a}\\
\noalign{\vskip 4pt}%
\frac{\delta \tilde{H}_{2}}{\delta b}
\end{array}\right),
\end{equation}
where 
\[
\tilde{H}_{0} = \int \mathrm{d}y\,(a-b),\quad \tilde{H}_{1} = \int
\mathrm{d}y\,ab,\quad \tilde{H}_{2} = \int
\mathrm{d}y\,\left(a^{2}b-ab^{2} + a_{y}b\right).
\] 
The other two Hamiltonian structures have been studied earlier to have
the forms
\begin{eqnarray}
\widetilde{\cal D}_1 & = &\left (\begin{array}{cc} -2a\partial -a_y & 
\partial^2 +  (a-b)\partial + a_y  \\
\noalign{\vskip 4pt}%
-\partial^2 +(a-b)\partial -b_y &
 2b\partial + b_y \end{array} \right ),\nonumber\\
\noalign{\vskip 4pt}%
\widetilde{\cal D}_2 & = & \left (\begin{array}{cc} 0 & \partial \\
\noalign{\vskip 4pt}%
 \partial & 0
\end{array} \right ). 
\end{eqnarray}
Similarly, we can write the system of equations (\ref{mkdv'}) as
\begin{equation}
\left(\begin{array}{c}a \\
\noalign{\vskip 4pt}%
 b \end{array}\right )_{\tau}=\widetilde{\cal D}_0 \left(\begin{array}{c}
\frac{\delta \tilde{H}_{1}}{\delta a}\\
\noalign{\vskip 4pt}%
\frac{\delta \tilde{H}_{1}}{\delta b}
\end{array}\right)  = 
\widetilde{\cal D}_1 \left(\begin{array}{c}
\frac{\delta \tilde{H}_{2}}{\delta a}\\
\noalign{\vskip 4pt}%
\frac{\delta \tilde{H}_{2}}{\delta b}
\end{array}\right) = \widetilde{\cal D}_2 \left(\begin{array}{c}
\frac{\delta \tilde{H}_{3}}{\delta a}\\
\noalign{\vskip 4pt}%
\frac{\delta \tilde{H}_{3}}{\delta b}
\end{array}\right),
\end{equation}
where  
\begin{equation}
\tilde{H}_3 = \frac{1}{2} \int \mathrm{d}y \big( 2b^3a - 3b^2a_y - 6b^2a
+2ba_{yy} +  6ba_ya + 2ba^3\big).
\end{equation} 
A careful analysis shows that the new Hamiltonian operator
$\widetilde{\cal D}_0$ can, in fact, be written as   
\begin{equation}
\widetilde{\cal D}_0 = - \widetilde{\cal D}_1 \widetilde{\cal
  D}_2^{-1} \widetilde{\cal D}_1. 
\end{equation} 
This shows that the three Hamiltonian structures are related through
the recursion operator
\[
{\cal R} = \widetilde{\cal D}_{1}\widetilde{\cal D}_{2}^{-1},
\]
as
\[
\widetilde{\cal D}_{1} = {\cal R} \widetilde{\cal D}_{2},\quad
\widetilde{\cal D}_{0} = - {\cal R}^{2} \widetilde{\cal D}_{2}.
\]
Consequently, these define compatible Hamiltonian structures and 
the associated Nijenhuis torsion tensor vanishes \cite{huang}.

To close this section we point out that the system of equations
(\ref{new'})  can be mapped to the modified dispersive water  
wave equations \cite{maciej} under the transformation 
\begin{equation}
 a \rightarrow -v,\quad  b \rightarrow w-v, \quad \tau
 \rightarrow -\frac{\tau}{2}. 
\end{equation}
Namely, under this transformation (\ref{new'}) goes into 
\begin{eqnarray}
v_{\tau} &=& \frac{1}{2} \big( -v_y + 2vw -v^2 \big)_y,\nonumber\\
\noalign{\vskip 4pt}%
w_{\tau} &=& \frac{1}{2} \big( w_x -2v_y -2v^2 +2vw +w^2 \big)_y,
\end{eqnarray} 
which corresponds to the modified dispersive water wave equation.
  
\section{Dispersionless Limit:}
Given a dispersive system, one can obtain the dispersionless limit by
taking the long wavelength limit \cite{zakharov}. 
In this limit the three component system obtained in (\ref{equation3})
takes the form 
\begin{eqnarray}
f_{\tau} &= & \Big ( - f^3 -3g^2f +3h^2f \Big )_y,\nonumber\\
\noalign{\vskip 4pt}%
g_{\tau} &=& \Big  ( - 5g^3-3gf^2+3h^2g  \Big )_y+ 6h_{y}hg,\nonumber
\\
\noalign{\vskip 4pt}%
h_{\tau} &=& \Big  ( + 5h^3-3hf^2-3hg^2 \Big )_y -6
hg_{y}g.\label{equation4} 
\end{eqnarray}
This can be written in the Hamiltonian form with the dispersionless
limit of the  Hamiltonian structure $\widetilde{\cal D}$ in
(\ref{structure}) 
\begin{eqnarray}
\widetilde{\cal D}^{(1,1)} &=&  - 2f_y\partial^{-1}f_y +\partial f^2 + f^2
\partial, \nonumber  \\
\noalign{\vskip 4pt}%
{\cal D}^{(1,2)} &=&  -2f_y\partial^{-1}g_y  + 2\partial gf +2g\partial f,
\nonumber  \\
\noalign{\vskip 4pt}%
\widetilde{\cal D}^{(1,3)} &=&  -2f_y\partial^{-1}h_y + 2\partial hf +
2h\partial f, \nonumber  \\
\noalign{\vskip 4pt}%
\widetilde{\cal D}^{(2,2)} &=& -2g_y\partial^{-1}g_y  + \partial(f^2+4g^2) +
(f^2+4g^2) \partial, \nonumber  \\
\noalign{\vskip 4pt}%
\widetilde{\cal D}^{(2,3)} &=& -2g_y\partial^{-1}h_y + 4(\partial hg +
hg \partial),\nonumber \\
\noalign{\vskip 4pt}%
\widetilde{\cal D}^{(3,3)} &=& -2h_y\partial^{-1}h_y +\partial (4h^2-f^2) +
(4h^2-f^2) \partial,  
\end{eqnarray}
and the Hamiltonian $H= \int \mathrm{d}y\left(- f^2 - g^2 + h^2\right)$.

If we now define the new variables 
\begin{equation}
F = {\frac{3}{4}} f^2 - {\frac{3}{4}} G, \quad G=-3g^2+ H, \quad
H=3h^2,\label{relation}
\end{equation}
then in these variables the dynamical equations can be written as
\begin{eqnarray}
F_{\tau} &= & \Big ( \big( F^2  - \frac{3}{4} G^2 - 2 GF\big )_y +G_yF \Big
),\nonumber  \\
\noalign{\vskip 4pt}%
G_{\tau} &=& \Big  ( \big ( G^2+2GF \big )_y +2GF_y\Big ), \nonumber
\\
\noalign{\vskip 4pt}%
H_{\tau} &=& \Big  ( \big ( 2HF+HG \big )_y +2HF_y\Big ).
\end{eqnarray}
The Hamiltonian structure correspondingly takes the form
\begin{eqnarray}
\widetilde{\cal D}^{(1,1)} &=&\partial(4F^2 - \frac{9}{2}G^2-15GF) + (4F^2 -
\frac{9}{2}G^2-15GF) \partial  -2 F_y\partial^{-1}F_y, \nonumber  \\
\noalign{\vskip 4pt}%
\widetilde{\cal D}^{(1,2)} &=& \partial(3G^2+20FG)+
(3G^2+20FG)\partial + 4FG_y 
+3G_yG - 2 F_y\partial^{-1}G_y, \nonumber\\
\noalign{\vskip 4pt}%
\widetilde{\cal D}^{(1,3)} &=& 20 \partial HF + (20HF+6HG)  \partial +
4FH_y 
+9HG_y -2 F_y\partial^{-1}H_y, \nonumber  \\
\noalign{\vskip 4pt}%
\widetilde{\cal D}^{(2,2)} &=& \partial (4G^2 - 16GF ) + (4G^2-16GF)
\partial  -2 G_y\partial^{-1}G_y, \nonumber  \\
\noalign{\vskip 4pt}%
\widetilde{\cal D}^{(2,3)} &=& - 16 \big( \partial HF+ HF \partial \big ) -
8HG\partial -4HG_y+12H_yG_y -2 G_y\partial^{-1}H_y,\nonumber  \\
\noalign{\vskip 4pt}%
\widetilde{\cal D}^{(3,3)} &=& \partial (16HF+12HG+16H^2) +(16HF+12HG+16H^2)
\partial\nonumber\\
\noalign{\vskip 4pt}%
& & \qquad  -2 H_y\partial^{-1}H_y.  
\end{eqnarray}
As we see, the first two equations do not depend on $H$ and the third
equation allows us to set $H=0$ consistently. In this case, with a
redefinition of variables $f = 3a-2b$ and $g=4(b-a) $, the first two
equations reduce to a much simpler system, namely, 
\begin{eqnarray}
a_{\tau} &=& \Big ( 5a^2-4ab \Big )_y,\nonumber \\
\noalign{\vskip 4pt}%
b_{\tau} &=& \Big ( - 2b^2+2ab \Big )_y +2ba_y.\label{new''}
\end{eqnarray}
In this case, the Dirac reduction of the Hamiltonian structure becomes
singular. Nevertheless, we have explicitly verified, using the method
of prolongation \cite{olver}, that the
Hamiltonian structure 
\begin{eqnarray}
\widetilde{\cal D}_0 ^{(1,1)} &=&  4 \partial (4a^2-3ab) + 4(4a^2-3ab)\partial
-2a_y\partial^{-1}a_y, \nonumber\\
\noalign{\vskip 4pt}%
\widetilde{\cal D}_0^{(1,2)} &=& 4\partial (2ab - b^2) + 4(2ab-b^2)
\partial - 4ba_y - 4b_yb -2a_y\partial^{-1}b_y,\nonumber\\
\noalign{\vskip 4pt}%
\widetilde{\cal D}_0^{(2,2)} &=& 4(\partial b^2 + b^2\partial
)-2b_y\partial^{-1}b_y, 
\end{eqnarray}
satisfies Jacobi identity and generates the dynamical equations as a
Hamiltonian system with the Hamiltonian $\tilde{H}_{0} = \frac{1}{3}
\int \mathrm{d}y\,a$. Furthermore, this new system of equations
(\ref{new''}) is, in fact, bi-Hamiltonian  
\begin{equation}
\left(\begin{array}{c}a \\
\noalign{\vskip 4pt}%
 b \end{array}\right )_{\tau}=\widetilde{\cal D}_0
\left(\begin{array}{c}
\frac{\delta \tilde{H}_{0}}{\delta a}\\
\noalign{\vskip 4pt}%
\frac{\delta \tilde{H}_{0}}{\delta b}\end{array}\right) = 
\widetilde{\cal D}_1 \left(\begin{array}{c}
\frac{\delta \tilde{H}_{1}}{\delta a}\\
\noalign{\vskip 4pt}%
\frac{\delta \tilde{H}_{1}}{\delta b}\end{array}\right),
\end{equation}
where $\tilde{H}_1= \int \mathrm{d}y\left(20a^2-16ab\right)$ and  
\begin{equation}
\widetilde{\cal D}_1=\frac{1}{12} \left( \begin{array}{cc}
\partial a+ a \partial & \partial b+ b \partial \\ 
\partial b+ b \partial & \partial b+ b \partial 
\end{array} \right ).
\end{equation}

We can now construct the recursion operator $R=\widetilde{\cal
  D}^{-1}\widetilde{\cal D}_0$ and obtain the infinite series of conserved  
charges. The first few have the explicit forms
\begin{eqnarray}
\tilde{H}_2 &=& \int \mathrm{d}y\, a(-21a^2-8b^2+28ab), \nonumber\\
\noalign{\vskip 4pt}%
\tilde{H}_3 &=& \int \mathrm{d}y\, a(-429a^3+729a^2b-432ab^2+64b^3),
\nonumber \\ 
\noalign{\vskip 4pt}%
\tilde{H}_4 &=& \int \mathrm{d}y\, a(-2431a^4 + 5720a^3b - 4576a^2b^2
+1408ab^3-128b^4).
\end{eqnarray}
In addition to these polynomial conserved charges, we can also
construct a nonpolynomial series of conserved charges using  
known techniques in  hydrodynamics, which we explain below. 

Let us note that
\begin{equation}
{\hat H}_1 = \int \mathrm{d}y\,{\sqrt { \frac{a}{b} -1 }}, \quad {\hat
  H}_2 = \int \mathrm{d}y\,a{\sqrt {a-b }}\label{charge1} 
\end{equation}
also define conserved quantities for the system. These charges have
been constructed using the analog of the  
Tricomi equation used in the theory of polytropic gas
dynamics \cite{pavlov}. Namely, if a conserved density $H$
($\hat{H}= \int \mathrm{d}y\,H$)  depends on $a$ and $b$
only, it satisfies $ H_{\tau} = G_y $ where $G$ depends only on $ a$ and
$b$. Using (\ref{new''}), this can be shown to lead to the following 
analog of the Tricomi equation 
\begin{equation}
4aH_{a,b} + 2aH_{aa} +2bH_{bb}+H_b=0,\label{tricomi}
\end{equation}
where the subscripts denote derivatives with respect to the particular
variable. 
The conserved quantities $H_i$ in (\ref{charge1}) can be obtained as
particular solutions of the Tricomi equation. 
We note that if we scale $H \rightarrow {\sqrt {(a-b)}} H $ then
equation (\ref{tricomi}) transforms to 
\begin{equation}
4aH_{a,b} + 2aH_{aa} +2bH_{bb} + 3H_b=0.\label{tricomi1}
\end{equation}
The general solutions of equations (\ref{tricomi}) and (\ref{tricomi1})
can be written as
\begin{equation}
H_n=\sum_{k=0}^{n-1}\lambda_{k,n}a^{n-k}b^k ,
\end{equation}
where 
\begin{equation}
\lambda_{k,n} = -2\lambda_{k-1,n}\frac{(n-k+1)(n-k)}{4k(n-k)+2k(k-1)+zk}
\end{equation}
for $ k > 0$ with $\lambda_{0,n} =0$ and $z=1$ for (\ref{tricomi})
while $z=3$ for (\ref{tricomi1}). In the first case, we have
polynomial charges the second leads to nonpolynomial charges (because
of the factor $\sqrt{a-b}$).

If we  further change the variable $b$ to $b= 3 c^2$, then equation
(\ref{new''}) has the conservative form 
\begin{eqnarray}
a_{\tau} & = & \Big ( 5a^2 -12ac^2 \Big )_y,\nonumber \\
\noalign{\vskip 4pt}%
c_{\tau} & = & \Big ( -4c^3 + 2ac \Big )_y,
\end{eqnarray}
which can be written in the bi-Hamiltonian form
\begin{equation} 
 \left(\begin{array}{c}a \\
\noalign{\vskip 4pt}%
 c \end{array}\right )_{\tau} = {\cal D}_0 \left(\begin{array}{c}
\frac{\delta H_{0}}{\delta a}\\
\noalign{\vskip 4pt}%
\frac{\delta H_{0}}{\delta b}
\end{array}\right) =  
{\cal D}_1 \left(\begin{array}{c}
\frac{\delta H_{1}}{\delta a}\\
\noalign{\vskip 4pt}%
\frac{\delta H_{1}}{\delta b}
\end{array}\right),
\end{equation}
where 
\begin{equation}
{\cal D}_1=\frac{1}{12} \left( \begin{array}{cc}
\partial a+ a \partial & c \partial \\ 
\partial c & \frac{1}{6} \partial  
\end{array} \right ),
\end{equation}
while ${\cal D}_0$ has the matrix elements
\begin{eqnarray}
{\cal D}_0^{(1,1)} &=&  4 \partial (4a^2-9ac^2) + 4(4a^2-9ac^2)\partial
-2a_y\partial^{-1}a_y,\nonumber\\
\noalign{\vskip 4pt}%
{\cal D}_0^{(1,2)} &=& -4(3c^3-2ca)\partial  + 2ca_y
-2a_y\partial^{-1}c_y,\nonumber  \\
\noalign{\vskip 4pt}%
{\cal D}_0(2,2) &=& \partial c^2 + c^2\partial  -2c_y\partial^{-1}c_y, 
\end{eqnarray}
and $H_0=\frac{1}{3} \int \mathrm{d}y\,a, H_1= \int
\mathrm{d}y\left(20a^2-48ac^2\right)$.

Finally let us comment on the different possible reductions of the system of
equations (\ref{equation4}). We note that when $ h=0 $, we have  
\begin{eqnarray}
f_{\tau} &= & \Big ( - f^3 -3g^2f  \Big )_y, \nonumber \\
\noalign{\vskip 4pt}%
g_{\tau} &=& \Big  ( - 5g^3-3gf^2 \Big )_y, 
\end{eqnarray}
and it corresponds to the dispersionless limit of equation
(\ref{mkdv}). We will not discuss this further.
If we set $ f=0 $, then (\ref{relation}) leads to
$ F =\frac{9}{4} (g^{2}-h^{2}), G=- 3(g^2 - h^{2}), H=3h^2$ and we
have the system of equations 
\begin{eqnarray}
G_{\tau} &=& -\frac{5}{4}  ( G^2 )_y, \nonumber \\
\noalign{\vskip 4pt}%
H_{\tau} &=&  -\frac{1}{2} ( HG )_y  + HG_y, 
\end{eqnarray}
where the first equation is decoupled.

\section{Conclusion:}

In this paper, we have constructed the two component supersymmetric
generalized Harry Dym equation which is integrable and have studied
various properties of this
model in the bosonic limit. In particular, we find a new integrable
model in the bosonic limit which under a hodograph transformation maps
to a system of three coupled MKdV equations. We have shown how the
Hamiltonian structure transforms under a hodograph transformation and
studied the properties of the system under a further reduction to a
two component system. We find a third Hamiltonian structure for this
system making this a genuinely tri-Hamiltonian system (it was known
earlier to be a bi-Hamiltonian system). We have clarified the
connection of 
this sytem to the modified dispersive water wave equation and have
studied 
various properties of our model in the dispersionless limit.

\section*{Acknowledgment:}

This work was supported in part by US DOE grant number
DE-FG-02-91ER40685 as well as NSF-INT-0089589.

\end{document}